\documentclass[12pt]{JHEP}
\usepackage{epsfig}

\font\blackboard=msbm10 at 12pt
\font\blackboards=msbm7
\font\blackboardss=msbm5
\newfam\black
\textfont\black=\blackboard
\scriptfont\black=\blackboards
\scriptscriptfont\black=\blackboardss

\newcommand{\NP}{{\rm Nucl.\ Phys.\ }}

\newcommand{\PL}{{\rm Phys.\ Lett.\ }}

\def\math@note#1{\gdef\@eqnlabel{LAB: #1}}
\title{A Perturbative Analysis of Tachyon Condensation}
\author{Washington Taylor\\
{Center for Theoretical Physics} \\
{MIT, Bldg.  6-308} \\
{Cambridge, MA 02139, U.S.A.} \\
{\tt wati@mit.edu}}

\abstract{Tachyon condensation in the open bosonic string is analyzed
using a perturbative expansion of the tachyon potential around the
unstable D25-brane vacuum.  Using the leading terms in the tachyon
potential, Pad\'e approximants can apparently give the energy of the
stable vacuum to arbitrarily good accuracy.  Level-truncation
approximations up to level 10 for the coefficients in the tachyon
potential are extrapolated to higher levels and used to find
approximants for the full potential.  At level 14 and above, the
resulting approximants give an energy less than -1 in units of the
D25-brane tension, in agreement with recent level-truncation results
by Gaiotto and Rastelli.  The extrapolated energy continues to
decrease below -1 until reaching a minimum near level 26, after which the
energy turns around and begins to approach -1 from below.  Within the
accuracy of this method, these results are completely consistent with
an energy which approaches -1 as the level of truncation is taken to
be arbitrarily large.}

\preprint{MIT-CTP-3298, hep-th/0208149}
\begin{document}

\baselineskip16pt
\parskip=4pt

\section{Background}
\label{sec:background}

Much effort has been expended over the last few years to confirm Sen's
conjectures \cite{Sen-universality} on tachyon condensation in the
open bosonic string (for reviews see \cite{reviews}).  Sen suggested
that the unstable open string vacuum in 26 dimensions should be
understood as containing a space-filling D25-brane.  Sen's first
conjecture states that the open bosonic string has a nontrivial
locally stable vacuum with energy $-T_{25} V$ relative to the unstable
perturbative open string vacuum, where $T_{25}$ is the tension of the
D25-brane, and $V$ is the volume of spacetime; this new stable vacuum
should correspond to the true closed string vacuum with no D-branes.
This conjecture has been tested by several authors using the approach
of truncating Witten's open string field theory \cite{Witten-SFT} to
include only fields below a certain total oscillator level.  In
\cite{ks-open} Kostelecky and Samuel used level truncation up to level
4, including 10 fields.  These authors found that at low levels of
truncation Witten's theory has a stable vacuum solution, and that the
energy of this stable vacuum seems to converge as the level of
truncation of the theory is increased.  At that time, however,
D-branes had not yet been understood, and there was no natural way to
interpret the energy of this new vacuum.  In \cite{Sen-Zwiebach} Sen
and Zwiebach reconsidered this calculation in the light of Sen's
conjectures.  They found that at level 4, the vacuum energy of the
nontrivial solution of the level-truncated theory is 0.986 times the
energy $-E_0 =-T_{25}V$ predicted by Sen.  In \cite{Moeller-Taylor},
Moeller and Taylor extended this calculation to higher levels.  We
found that as the level of truncation is increased, the vacuum energy
continues to decrease, until at level 10 the vacuum energy reaches
$-0.9991\, E_0$.  These calculations seemed to indicate that level
truncation gives a systematic approximation to the full string field
theory, and that the vacuum energy computed in level truncation would
converge monotonically to $-E_0$.  The accuracy of this picture was
thrown into doubt when Gaiotto and Rastelli recently announced
\cite{Gaiotto-Rastelli} that when the level of truncation is increased
further, the vacuum energy {\it overshoots} $-E_0$.  They found that
at level 14 the vacuum energy becomes $-1.0002\, E_0$, and that the
energy continues to decrease monotonically until at level 18 the
vacuum energy is $-1.0005\, E_0$.

In this paper we use an alternative approach based on perturbation
theory to estimate the energy of the nontrivial vacuum of open string
field theory.  We use level truncation to determine the values of the
coefficients in a perturbative expansion of the effective
zero-momentum tachyon potential around the usual open string vacuum.
We then extrapolate these values to higher levels, and use the actual and
predicted values of the coefficients at various levels to compute Pad\'e
approximants for the effective potential.  The resulting Pad\'e
approximants seem to give highly reliable estimates for the vacuum
energy.  Furthermore, the Pad\'e approximants based on the extrapolated
perturbative coefficients give rise to energies which closely match
those found by Gaiotto and Rastelli.  After overshooting the desired
value of $-E_0$ for the energy, however, the predicted energies turn
around again near level 26, eventually returning close to the desired
value.  These calculations show that the results of Gaiotto and Rastelli
are fully compatible both with earlier computations of the
coefficients of the effective tachyon potential and with the
conjecture that the energy of the nontrivial vacuum in level-truncated
SFT asymptotically approaches the value predicted by Sen as the level
is taken arbitrarily large.

In this paper we work exclusively in Feynman-Siegel gauge.  Other
choices of gauge were explored in \cite{Ellwood-Taylor-gauge}; for
some other gauge choices, the level-truncated approximations to the
energy were found to go below $-E_0$ (sometimes at levels as low as $L
= 4$), and to be non-monotonic.  The results of this paper suggest
that this is in fact the generic behavior of level-truncated
approximations in any gauge; we expect that applying the methods of
this paper to the calculation in other gauges will give similar
results.

In Section 2 we discuss the calculation of coefficients in the
effective tachyon potential.  In Section 3 we describe the method of
Pad\'e approximants and their use in studying the vacuum energy of SFT.
In Section 4 we use Pad\'e approximants on the coefficients computed in
Section 2 and summarize our results.

\section{Perturbative tachyon potential}
\label{sec:perturbative}

The effective potential for the $p = 0$ tachyon field
$\phi$ is given by
\begin{eqnarray}
V (\phi)  & = & \sum_{n = 2}^{ \infty}  c_n (\kappa g)^{n-2} \phi^n
\label{eq:v}\\
& = & -\frac{1}{2}\phi^2 + (\kappa g) \phi^3 + c_4 (\kappa g)^2
\phi^4 + c_5 (\kappa g)^3 \phi^5 +\cdots\nonumber
\end{eqnarray}
where $\kappa=3^{7/2}/2^7 \approx 0.365$  and where
$c_4, c_5, \ldots$ are numerical constants.  Closed form expressions
for these coefficients were given in
\cite{ks-exact,Samuel-kn,WT-perturbative}.  These expressions cannot
be evaluated exactly, however, and must be approximated numerically.
Level truncation of Witten's string field theory gives an efficient
means of approximating these coefficients
\cite{ks-open,Moeller-Taylor,WT-perturbative}.
In \cite{Moeller-Taylor}, we computed the coefficients
up to $c_{60}$ at truncation levels up to (10, 20), meaning that all
fields up to level 10 and interactions up to total level 20 were
included.

For example, for $c_4$ successive level-truncation approximations give
\begin{eqnarray}
c_4^{[2]} & = & -\frac{34}{27} \approx -1.259259\\
c_4^{[4]} &  \approx & -1.472489 \nonumber\\
c_4^{[6]} &\approx & -1.556198 \nonumber\\
c_4^{[8]} &  \approx& -1.600425 \nonumber\\
c_4^{[10]} & \approx& -1.627694 \nonumber
\end{eqnarray}
where by $c_4^{[L]}$ we denote the approximation to $c_4$ in level
approximation $(L, 2L)$.  Note that throughout this paper we
systematically use level truncations $(L, 2L)$;  empirical
observations indicate that level $(L, 2L)$ approximations are
generally very close to level $(L, 3L)$ approximations, although
the former are easier to compute
\cite{Moeller-Taylor,Gaiotto-Rastelli}.   An analytic expression for
$c_4$ given in \cite{ks-exact} was numerically evaluated in
\cite{ks-open,WT-perturbative} and found to give 
\begin{equation}
c_4 \approx -1.742
\pm 0.001\,.
\label{eq:c4-mt}
\end{equation}
In \cite{WT-perturbative}, approximations to $c_4$ were computed using
oscillator level truncation of an exact expression written in terms
of an infinite matrix of Neumann coefficients.  These approximations
$c_4^{(L)}$ were computed up to oscillator level $L = 100$, and were
found to behave at large $L$ as
\begin{equation}
c_4^{(L)} \approx -1.742 + 0.80 \;L^{-1} +{\cal O} (L^{-2})\,.
\label{eq:c4-oa}
\end{equation}

To use the Pad\'e approximant method described in the next section, we
need an accurate approximation to the coefficients $c_{n}$ for $n$ up
to $26$ or so.  While the method of \cite{WT-perturbative} is much more
efficient for computing $c_n$ for small $n$ at high levels, at large
values of $n$ this approach becomes more difficult as the number of
diagrams which must be summed increases quickly with $n$.  Thus, in this paper
we will use the approximate coefficients $c_n^{[L]}$ computed in
\cite{Moeller-Taylor} up to level $L = 10$.  From the analysis of
\cite{WT-perturbative}, however, we expect that these approximate
coefficients will have corrections which can be expressed as a power
series in $1/L$.  This allows us to use the data for $L \leq 10$ to
predict the values of the approximate coefficients for $L > 10$.  For
example, using the values of $c_4^{[6]}, c_4^{[8]}, c_4^{[10]}$ from
(\ref{eq:c4-mt}) we can fit to a function of the form $a + b/L + c/L^2$ to get
\begin{equation}
c_4^{\{2, L\}} \approx-1.74226+0.59482 \, L^{-1}-0.10987\, L^{-2} \,.
\label{eq:c4-2}
\end{equation}
Using the value of $c_4^{[4]}$ in addition, and allowing a term of
order $L^{-3}$ gives
\begin{equation}
c_4^{\{3, L\}} \approx-1.74204+0.59213\, L^{-1}-0.09933 \,L^{-2} 
-0.01346 \, L^{-3}\,.
\label{eq:c4-3}
\end{equation}
Including $c_4^{[2]}$ and a term of order $L^{-4}$ gives
\begin{equation}
c_4^{\{4, L\}} \approx-1.74190+0.59013\, L^{-1}-0.08916 \,L^{-2} 
-0.03550 \, L^{-3} + 0.01718\, L^{-4}\,.
\label{eq:c4-4}
\end{equation}
The low order terms in these extrapolations seem to be converging
well, so we expect that (\ref{eq:c4-2}-\ref{eq:c4-4}) give
increasingly good estimates of the level-truncated approximations to
the coefficient $c_4$.  We have performed similar extrapolations for
the higher coefficients $c_n$ with similar results.  We will use these
extrapolations in Section 4 to estimate the value of the vacuum energy
in various level-truncated approximations.

\section{Pad\'e approximants}
\label{sec:Pade}

The power series expansion (\ref{eq:v}) of the tachyon effective
potential $V (\phi)$ has a finite radius of convergence of
approximately $\phi_r \approx 0.25/g$ \cite{Moeller-Taylor}.  This
finite radius of convergence arises due to the breakdown of
Feynman-Siegel gauge when $\phi$ approaches the critical value $\phi_c
\approx -0.25/g$ \cite{Ellwood-Taylor-gauge}.  At this point the
function $V (\phi)$ has a square root branch cut singularity.  The
nontrivial vacuum occurs near $\phi \approx 1/g$, so that the series
expansion of $V (\phi)$ is badly divergent near the true vacuum.
Thus, we cannot directly study the local minimum of the tachyon
effective potential by summing the perturbation series (\ref{eq:v}).

A powerful method for dealing with series expansions outside their
radius of convergence is given by the method of Pad\'e
approximants.  Given a power series
\begin{equation}
f (x) = a_0 + a_1 x + a_2x^2 + a_3x^3 + \cdots,
\label{eq:fx}
\end{equation}
the Pad\'e approximant $P^M_N (x)$ is the unique rational function
with numerator of degree $M$ and denominator of degree $N$ which
agrees with the power series expansion (\ref{eq:fx}) out to order $M +
N$
\begin{equation}
P^M_N (x) = \frac{\sum_{ i = 0}^{M} b_i x^i}{ \sum_{j = 0}^{N} d_j
  x^j}  = \sum_{k = 0}^{ N + M}  a_k x^k +{\cal O} (x^{N + M + 1})\,.
\end{equation}

While the partial sums computed by truncating (\ref{eq:v}) at finite
order do not converge near the stable tachyon vacuum, these partial
sums may be used to compute Pad\'e approximants with much better
convergence behavior.  For example,
consider the approximate perturbative potential
given by truncating (\ref{eq:v}) at $\phi^K$ and using the level $2$
approximations $c_n^{[2]}$ to the coefficients.  
For $K = 4$, the resulting quartic 
\begin{equation}
-\frac{1}{2}\phi^2 + \kappa g \phi^3-\frac{34}{27}  (\kappa g) \phi^4
\end{equation}
has no local minima, while the Pad\'e approximant
\begin{equation}
P^3_1 (\phi) =\frac{-\frac{1}{2}\phi^2 + \frac{10}{27} \kappa g \phi^3
}{1 + \frac{34}{27}  \kappa g \phi} 
\end{equation}
does.

\FIGURE{
\epsfig{file=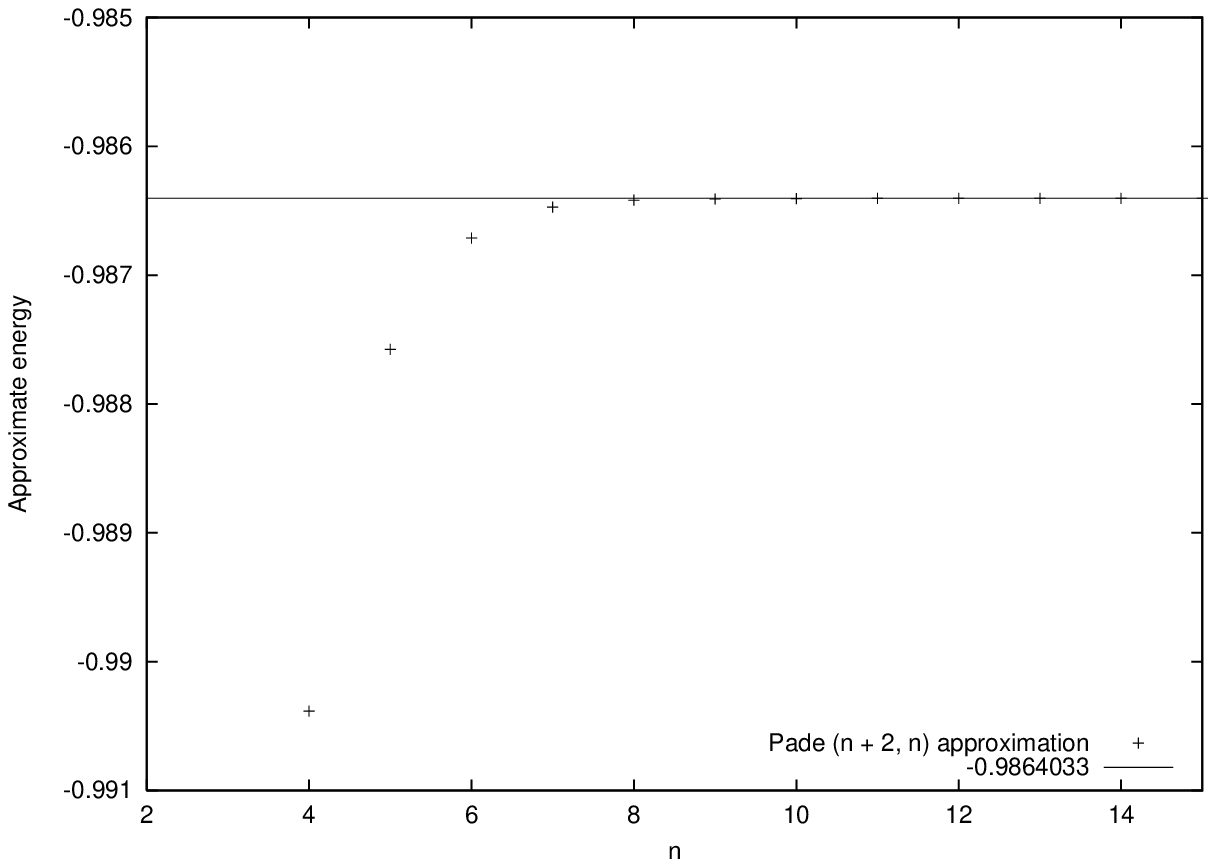,width=15cm}
\caption{\footnotesize Level 4 vacuum energy using Pad\'e approximants}
\label{f:Pade-4}
}

We have investigated the behavior of various families of Pad\'e
approximants using the coefficients of the tachyon effective potential
computed at low truncation levels.  We find that the family of
approximants $P^{n + 2}_n$ does very well at reproducing the effective
potential out to the nontrivial vacuum.  For example, at level 4 the
effective potential has a minimum with energy $-0.986403\, E_0$
\cite{Sen-Zwiebach,Moeller-Taylor}.  Using Pad\'e approximants on the level 4
coefficients we find that  the Pad\'e approximants $P^{n +
2}_n$ give increasingly accurate estimates of vacuum energy, giving
for example
(in units of $E_0$)
\begin{eqnarray}
E^{12}_{ 10} & = &  -0.9864053 \nonumber\\
E^{13}_{11} & = &  -0.9864036\\
E^{14}_{12} & = & -0.9864033 \nonumber
\end{eqnarray}
The Pad\'e approximants converge up to the limit of numerical precision
possible from the coefficients we have used.  The energies arising from
using Pad\'e approximants $P^{n + 2}_n$ to estimate the vacuum energy at
level 4 are graphed in Figure~\ref{f:Pade-4}.  Performing a similar
analysis on the approximate coefficients at other levels of
truncation, we find that up to level 10, the Pad\'e approximant
$P^{14}_{12}$ comes within $10^{-6}$ of the exact value of the energy
at the minimum of the approximate effective potential.  We expect
similar behavior at higher levels.  We use this approximant in
our further analysis.

\section{Estimating the vacuum energy}
\label{sec:energy}

In Section 2 we described a method for estimating the coefficients
$c_n^{[L]}$ which would be computed using level truncation at levels
$L > 10$.  In section 3 we described a method for using coefficients
in the effective tachyon potential to estimate the vacuum energy.  We
can now combine these approaches to estimate the vacuum energy which
would be computed in a level $L$ truncation with $L > 10$.

\FIGURE{
\epsfig{file=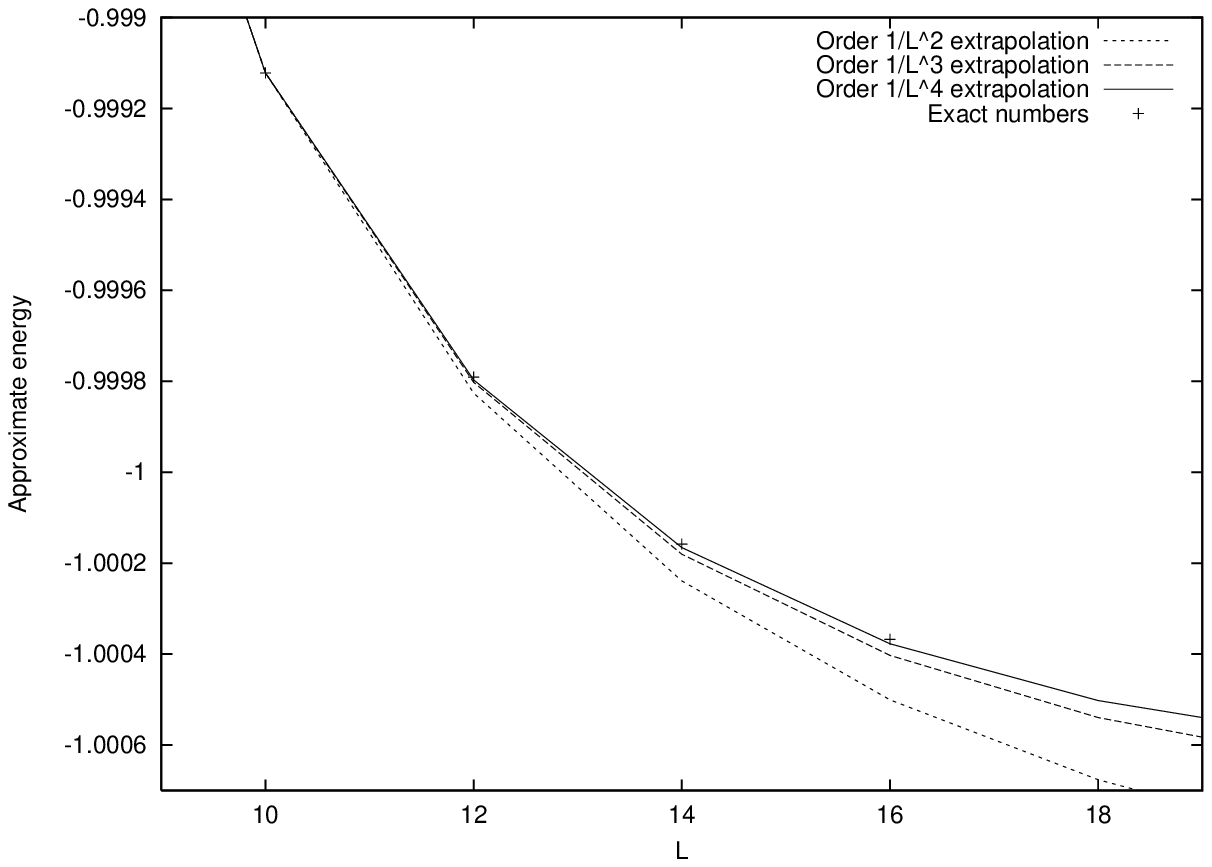,width=15cm}
\caption{\footnotesize Extrapolating vacuum energy in level truncation
$L > 10$}
\label{f:extrapolate-L}
}

We have used the Pad\'e approximant $P^{14}_{12}$ to estimate the vacuum
energy which would arise from the tachyon potential given by each of
the three families of extrapolated coefficients $c_n^{\{m, L\}}$ given in
(\ref{eq:c4-2}-\ref{eq:c4-4}) for $m = 2, 3, 4$.  The resulting vacuum
energies are graphed in Figure~\ref{f:extrapolate-L} and compared to the
exact results of Gaiotto and Rastelli at levels 12-16
\cite{Gaiotto-Rastelli}.  As can be seen from the graph, while the
order $1/L^2$ extrapolation of the perturbative coefficients gives an
energy which overshoots the correct values, the order $1/L^4$
extrapolation is very close to the exact values.  For example, at
level 16 Gaiotto and Rastelli found
\begin{equation}
E^{[16]} = -1.0003678.
\end{equation}
The order $1/L^4$ extrapolation gives
\begin{equation}
E^{\{4, 16\}} \approx -1.0003773
\end{equation}
while the order $1/L^3$ and $1/L^2$ extrapolations give $E^{\{3, 16\}}
\approx -1.0004030$ and $E^{\{2, 16\}} \approx -1.0005007$
respectively.

Seeing that we can fairly accurately reproduce the values found
by Gaiotto and Rastelli for the energy in the level-truncated theory
at levels $L > 10$, it is of great interest to extrapolate to still
higher values of $L$.  We have done this for each of the three
extrapolations we are using.  In each case, we find that at a
particular level the extrapolated energy reaches a minimum value and
then starts to climb back towards -1.   Table~\ref{t:energies} summarizes
for each of the extrapolations we are using the level at which the
extrapolated energy reaches a minimum, the energy at the corresponding
minimum, and the energy as $L \rightarrow \infty$ \TABLE{
\begin{tabular}{ || c  || c | c | c ||}
\hline
\hline
Order &  $L$ at minimum $E$  & $E_{{\rm min}}$ & $E_{\infty} $\\
\hline
\hline
$1/L^2 $ & 138 &-1.001247 & -1.001239\\
\hline
$1/L^3 $ & 34 &-1.000761 & -1.000444\\
\hline
$1/L^4 $ & 28 &-1.000661 & -1.000159\\
\hline
\hline
\end{tabular}

\caption{\footnotesize Minimum and asymptotic energies for different
  extrapolations}
\label{t:energies}
}
The extrapolated energy is graphed as a function of $\ln L$ in
Figure~\ref{f:log-energy}.

\FIGURE{
\epsfig{file=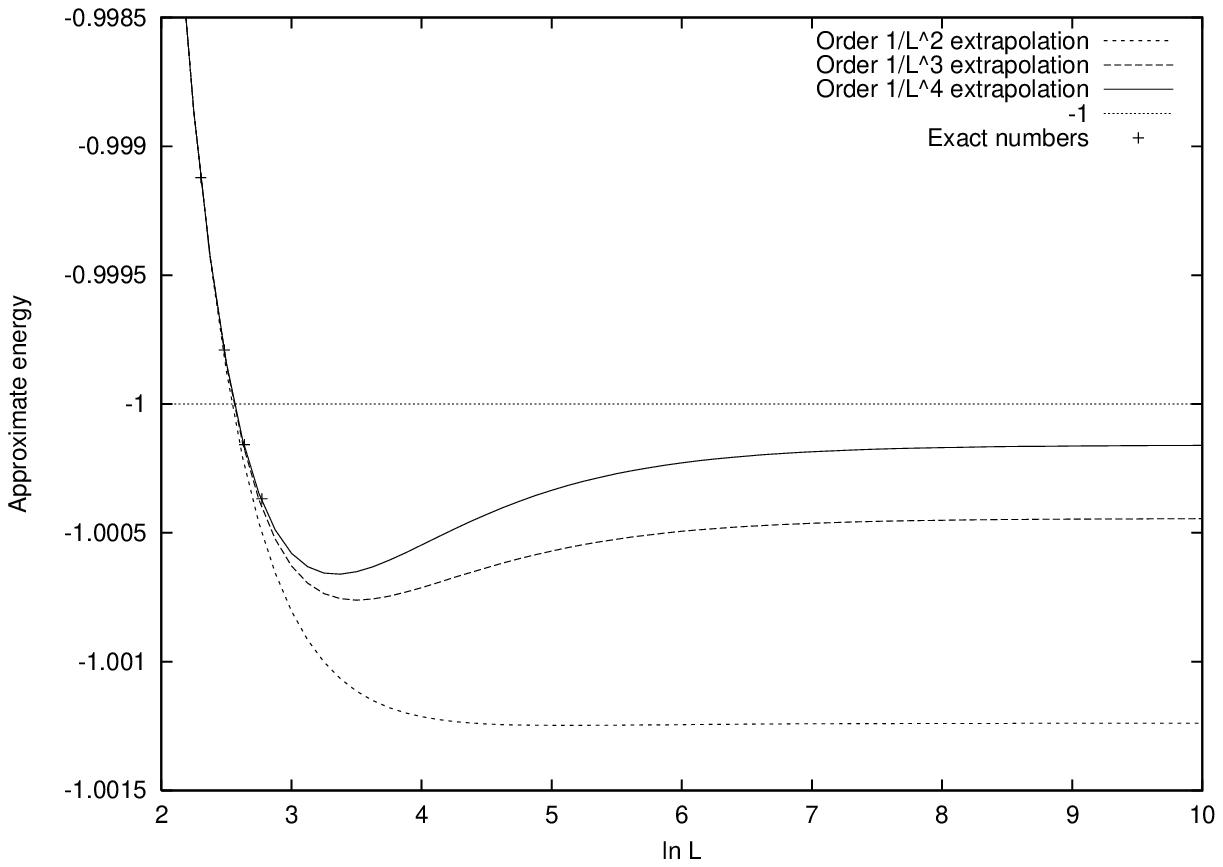,width=15cm}
\caption{\footnotesize Extrapolated vacuum energy for large $L$}
\label{f:log-energy}
}

It is clear that these results are compatible both with the
observation of Gaiotto and Rastelli that the energy shoots below -1 at
level 14 and with the conjecture that as $L \rightarrow \infty$ the
energy approaches $-1$ in a controlled fashion.  It would be nice to
have a clearer theoretical understanding of why the individual
coefficients in the perturbative expansion of the tachyon effective
potential have corrections described by a power series in $1/L$ in
level truncation, as this feature was crucial in getting the
computations in this paper to work out.

\section*{Acknowledgements}

I would like to thank Leonardo Rastelli, Ashoke Sen, and Ilya Sigalov
for helpful discussions.  I would also like to thank the Aspen Center
for Physics, where this work was done.  The numerical computations
described in this work were done using {\it Mathematica}.  This work
was supported by the DOE through contract \#DE-FC02-94ER40818.


\normalsize

\bibliographystyle{plain}

\begin{thebibliography}{10}

\bibitem{Sen-universality} 
A.~Sen,
``Universality of the tachyon potential,''
JHEP {\bf 9912}, 027 (1999),
{\tt hep-th/9911116}.


\bibitem{reviews}
K.~Ohmori,
``A review on tachyon condensation in open string field theories,''
{\tt hep-th/0102085};
P.~J.~De Smet,
``Tachyon condensation: Calculations in string field theory,''
{\tt hep-th/0109182};
I.~Y.~Arefeva, D.~M.~Belov, A.~A.~Giryavets, A.~S.~Koshelev and P.~B.~Medvedev,
``Noncommutative field theories and (super)string field theories,''
{\tt hep-th/0111208};
W.\ Taylor and B.\ Zwiebach, ``2001 TASI lectures'', {\it to appear}.

\bibitem{Witten-SFT}
E.\ Witten, ``Non-commutative geometry and string field theory,'' \NP {\bf
  B268} 253 (1986).

\bibitem{ks-open}
V.\ A.\ Kostelecky and S.\ Samuel, ``On a nonperturbative vacuum for the open
  bosonic string,'' \NP {\bf B336} 263 (1990).

\bibitem{Sen-Zwiebach}
A.~Sen and B.~Zwiebach,
``Tachyon condensation in string field theory,''
JHEP {\bf 0003}, 002 (2000),
{\tt hep-th/9912249}.



\bibitem{Moeller-Taylor}
N.~Moeller and W.~Taylor,
``Level truncation and the tachyon in open bosonic string field theory,''
Nucl.\ Phys.\ B {\bf 583}, 105 (2000),
{\tt hep-th/0002237}.

\bibitem{Gaiotto-Rastelli}
D.\ Gaiotto and L.\ Rastelli,
``Progress in open string field theory,''
Presentation by L.\ Rastelli at Strings 2002, Cambridge, England;
{\tt http://www.damtp.cam.ac.uk/strings02/avt/rastelli/}.

\bibitem{Ellwood-Taylor-gauge} 
I.\ Ellwood and W.\ Taylor,
``Gauge invariance and tachyon condensation in open string field
theory,''
{\tt hep-th/0105156}


\bibitem{ks-exact}
V.\ A.\ Kostelecky and S.\ Samuel, ``The static tachyon potential in the open
  bosonic string,'' \PL {\bf B207} 169 (1988).

\bibitem{Samuel-kn}
R.\ Bluhm and S.\ Samuel, ``The off-shell Koba-Nielsen formula,'' \NP
{\bf B323} 337 (1989). 

\bibitem{WT-perturbative}
W.~Taylor,
``Perturbative diagrams in string field theory,''
{\tt hep-th/0207132}.


\end{thebibliography}

\end{document}